\def\EqnRef#1{Eq.~\ref{#1}}
\def\FigRef#1{Fig.~\ref{#1}}

\def\JournalRef#1#2#3#4{#1 #2 (#4) #3}
\def\PR{Phys.Rev.}
\def\PRL{Phys.Rev.Lett.}
\def\EPJ{Euro.Phys.J.}
\def\NPBsupp{Nucl.Phys.Proc.Suppl.}

\def\etal{{\it et al.}}
\def\FullStop{.}

\def\DtoK{D\to Ke\nu}
\def\Dtopi{D\to \pi e\nu}
\def\ssbar{{s\overline{s}}}
\def\Dstoss{D_s\to \pi_\ssbar e\nu}

\def\Vec#1{{\mathbf #1}}
\def\ME{{\cal M}_\mu}
\def\GeV{{\rm GeV}}
\def\Vratio{|V_{cd}|/|V_{cs}|}
\def\Pcut{|\Vec{p}|^{cut}}

\def\BorderBox#1#2{\vbox{\hrule height #1%
                         \hbox{\vrule width #1%
                               #2%
                               \vrule width #1%
                              }%
                         \hrule height #1%
                        }%
                  }

\def\FigFudge{\vskip-1.0em}

\def\mQscalingFig{%
\begin{figure}[t]
\BorderBox{0pt}{%
\epsfig{
width=\columnwidth,
bbllx=18,bblly=22,bburx=596,bbury=380,
file=proc_ME_vs_1_over_mQ.ps}
}
\caption{%
Comparison of $\ME^\ssbar(\Vec{p}_\ssbar,\Vec{0})$  for  the  decay of
$B_s$ and $D_s$ mesons at $\beta=5.7$.}
\FigFudge
\label{fig:mQscaling}
\end{figure}
}

\def\DecayRateFig{%
\begin{figure}[t]
\BorderBox{0pt}{%
\epsfig{
width=\columnwidth,
bbllx=20,bblly=20,bburx=598,bbury=380,
file=b5.7_Combined_dGamma_vs_pf.ps}
}
\caption{%
Decay  rates   (with  statistical  errors)     versus daughter  recoil
momentum.}
\FigFudge
\label{fig:DecayRate}
\end{figure}
}

\def\ErrorTable{%
\begin{table}[hbt]
\caption{Summary of errors in $\Vratio$ at $\beta=5.7$}
\label{tab:Errors}
\begin{tabular}{lr}
\hline
source					&$\%$	\\ \hline
M.C. statistics				&$18$	\\
fits, excited state contamination	&$~2$	\\
residual cutoff dependence		&$<10$	\\
$a$ determination			&$~1$	\\
bare $m_s$ det., extrapolation to $m_l$	&$10$	\\
bare $m_c$ determination		&$~1$	\\ \hline
\end{tabular}
\end{table}
}

\input hyperbasics.tex

\documentstyle[twoside,fleqn,espcrc2,epsfig]{article}

\title{
\vskip-2.0ex\hbox to 6.25in {{\normalsize \hfil FERMILAB-Conf-98/333-T}}
Estimated Errors in $\Vratio$ from Semileptonic $D$ Decays}

\author{%
J.N. Simone%
\address{Fermilab, P.O. Box 500, Batavia, IL 60510, USA}%
\thanks{Presenter},
A.X. El-Khadra%
\address{Department of Physics,
University of Illinois, Urbana, IL 61801, USA},
S. Hashimoto${\rm^a}$%
\thanks{permanent address: KEK, Tsukuba, Japan},
A.S. Kronfeld${\rm^a}$,
P.B. Mackenzie${\rm^a}$
and
S.M. Ryan${\rm^a}$
}

\begin{document}

\begin{abstract}
We estimate statistical and systematic errors in the extraction of the
CKM ratio $\Vratio$ from exclusive $D$-meson semileptonic decays using
lattice QCD and anticipated new experimental results.
\end{abstract}

\maketitle

\section{INTRODUCTION}

High statistics  experimental  studies  of  $D$ mesons, such   as E831
(FOCUS)  and  E791, are  expected   to  yield ${\cal O}(10^6)$   fully
reconstructed  $D$  decays   with   better  momentum  resolution  than
preceding  experiments\cite{FOCUS,E791}.   Anticipating high precision
experimental results, we consider how well the CKM ratio $\Vratio$ can
be determined from  a  ratio of $\Dtopi$ to  $\DtoK$  decay  rates and
lattice QCD.

For the  Cabibbo allowed  $\DtoK$  decay, lower statistics  and poorer
resolutions in  earlier   experiments meant hadronic matrix   elements
could be described by functions of only two parameters: the CKM matrix
element times  a form factor  at zero momentum  transfer and  the pole
mass in a pole  dominance {\em Ansatz}.  The  PDG quotes a $3\%$ error
for $|V_{cs}|f_+^{K}(0)$  based upon such techniques\cite{PDG98}.  The
uncertainty due to this  {\it Ansatz} is  difficult to quantify.  They
quote  a   $14\%$ uncertainty  in  the  theoretical   determination of
$f_+^{K}(0)$ resulting in a $15\%$ uncertainty for $|V_{cs}|$.

$|V_{cd}|$  is   determined   with      an  error of      $7\%$   from
$\nu$-$\overline{\nu}$  production  of  charm   off   of  valence  $d$
quarks\cite{PDG98}. Combining this uncertainty and that for $|V_{cs}|$
we deduce  an  error of about   $17\%$ in the   current best value  of
$\Vratio$.

We examine uncertainties   in  determining $\Vratio$ directly from   a
ratio of semileptonic decay  rates.  We expect that both  experimental
and theoretical systematic errors common to  $\DtoK$ and $\Dtopi$ will
be reduced in ratios.

\section{PROCEDURE}
\label{sect:Procedure}

We suggest the ratio  of CKM matrix elements  can  be determined by  a
double ratio
\begin{equation}
\frac{|V_{cd}|}{|V_{cs}|}=
\frac{E_{\pi/K}(\Pcut)}{T_{\pi/K}(\Pcut)}\;\FullStop
\label{eqn:ExptOverTheory}
\end{equation}
The maximum recoil momentum cut, $\Pcut$, will be adjusted to minimize
combined experimental  and  theoretical errors\cite{LAT95}.  The ratio
of partial widths
\begin{equation}
E_{\pi/K}^2(\Pcut)
\equiv
\frac{
\int_0^{\Pcut}d\Gamma(\Dtopi)
}{
\int_0^{\Pcut}d\Gamma(\DtoK)
}
\label{eqn:GammaRatio}
\end{equation}
is to be determined experimentally.  $T_{\pi/K}(\Pcut)$ is provided by
theory calculation.  In the $D$-meson rest frame,
\begin{equation}
T_{\pi/K}^2(\Pcut)\equiv
\frac{
\int_{0}^{\Pcut}d|\Vec{p}|_\pi \Vec{p}_\pi^4 |f^\pi_+(E_\pi)|^2 /E_\pi
}{
\int_{0}^{\Pcut}d|\Vec{p}|_K \Vec{p}_K^4 |f^K_+(E_K)|^2 /E_K
}
\label{eqn:ThyRatio}
\end{equation}
where   hadronic matrix elements  are  expressed  conveniently by form
factors $f_+^X$.

Matrix elements
\begin{equation}
\ME^X(p_X,p_D)=\frac{\langle X(p_X)|V_\mu|D(p_D)\rangle}{\sqrt{2E_D2E_X}}
\label{eqn:MatrixElement}
\end{equation}
are determined in the continuum limit of our lattice results.  $\ME^X$
do  not depend upon the heavy  quark mass in  the infinite mass limit.
The {\em continuum} relation
\begin{eqnarray}
\ME^X(p_X,p_D)=\left[(p_D+p_X)_\mu f_+^X(q^2)  + \right.\qquad&& \nonumber \\
 \left. (p_D-p_X)_\mu f_-^X(q^2)\right]/\sqrt{2E_D2E_X}&&
\label{eqn:FormFactorDecomp}
\end{eqnarray}
is used to extract the form factors.

\section{MATRIX ELEMENTS}

We use  the tadpole-improved  Sheikholeslami-Wohlert quark action with
the   plaquette-determined  mean field  coefficient  $u_0$.  We obtain
${\cal     O}(a)$-improved    matrix  elements   using   the  Fermilab
interpretation for heavy  quarks and tree-level ${\cal O}(a)$-improved
currents\cite{Massive}.   Note that the vector current renormalization
factor cancels in ratio $T_{\pi/K}$.

We have   computed three-point correlators  at  $\beta=5.7$, $5.9$ and
$6.1$.  We determine the lattice   spacing, $a$, using the  charmonium
$1P$-$1S$  splitting and the bare   charm mass using the spin-averaged
$1S$ charmonium kinetic   mass.  The  same  gauge-field ensembles  and
procedures  were    used to    study   $B$   and   $D$  meson    decay
constants\cite{fBpaper}.     Reference~\cite{Btopi}   describes    our
$B\to\pi e\nu$ results on these lattices and  provides more details on
our three-point function calculations.

\DecayRateFig

For illustration, and to reduce  some known systematic errors, we take
$\Pcut=0.7\,\GeV$.   For  this preliminary   study we  take  our  best
estimate for  matrix  elements from  $\beta=5.7$.   We approximate the
$\DtoK$ decay by $\Dstoss$ where a pseudoscalar ``$\pi_\ssbar$'' meson
consists of  $s$  and  $\overline{s}$ valence quarks.   We  denote the
ratio     in   \EqnRef{eqn:ThyRatio}    for   the   $D_s$  decay    by
$T_{\pi/\ssbar}$.  The analysis for $\DtoK$ is underway.

In   \FigRef{fig:DecayRate} we    show differential decay    rates for
$\beta=5.7$.   The ratio      of areas  for  $\Pcut=0.7\,\GeV$   gives
$T_{\pi/\ssbar}=1.53\pm0.27$      where  the   error  is  statistical.
Systematic errors are the subject  of the rest  of the paper.

We use all three lattices to study the  cutoff dependence of $\Dstoss$
matrix elements.  We compare partial widths  at $\beta=5.7$ to partial
widths computed from  $\ME^\ssbar$   extrapolated  to the    continuum
linearly   with      the   lattice  spacing.      Matrix      elements
$\ME^\ssbar(\Vec{p}, \Vec{0})$  for a given  $\Vec{p}$ on each lattice
were found  by interpolating among  $\ME^\ssbar$ values calculated for
discrete  values  of $\Vec{p}$.  Discretization  errors   in the width
increase  with   $\Vec{p}$,   as   expected, reaching      $21\%$  for
$|\Vec{p}|=0.7\,\GeV$.  Momentum dependent  errors of  this  magnitude
were anticipated by errors  estimates for free-field tree-level  quark
matrix  elements\cite{LAT95}.  Errors for  the pion width are expected
to be similar to those for the  $\ssbar$ width. These errors, however,
will tend   to cancel in  ratios.  Hence  we  estimate residual cutoff
errors are below $10\%$ for $T_{\pi/\ssbar}(0.7)$.

\mQscalingFig

Our   bare charm masses were determined   in  quarkonia.  Charm masses
determined using   heavy-light mesons   differ due  to  discretization
errors and  the quenched approximation.   We  find  at  most a  $10\%$
difference  in     bare    $m_c$   comparing   the     two    methods.
Figure~\ref{fig:mQscaling},              which                compares
$\ME^\ssbar(\Vec{p},\Vec{0})$  calculated for  $D_s$ and $B_s$ mesons,
shows that $1/m_Q$  corrections are  small in this   case\cite{LAT95}.
Numerical results and pole dominance arguments indicate that $\ME^\pi$
has somewhat more $1/m_Q$ dependence  than $\ME^\ssbar$.  We  estimate
the  uncertainty  in adjusting   $m_c$  leads  to  a  $1\%$  error  in
$T_{\pi/\ssbar}$.

The quenched approximation leads  to systematic differences in lattice
spacing determinations.   We take   the charmonium   $1P$-$1S$ lattice
spacing  as  our best estimate  of   $a$.  Repeating  our analysis for
another reasonable  choice     for the   lattice  spacing,   $a(f_K)$,
determined by the  kaon  decay constant, changes  $T_{\pi/\ssbar}$  by
about  $1\%$.  This almost certainly  underestimates the  error due to
quenching.

Correlators are computed for light quark masses between about $0.4$ to
$1.2$  times  $m_s$.  For lighter   masses, correlators begin  to show
increasing  fluctuations caused by  nearby  unphysical  Dirac operator
zero-modes found on a small subset of our gauge configurations.  Hence
we must obtain matrix  elements for up  and down quarks by  ``chiral''
extrapolations of  matrix elements  computed for  masses close to  the
strange   quark.  Near zero  recoil  momentum these extrapolations are
more difficult  because matrix elements  are  sensitive to  the  light
quark mass.  This can be understood as the effect of nearby poles in a
pole dominance picture.   The effect  upon $T_{\pi/\ssbar}$ is  small,
however,  since  the decay  rate is  kinematically suppressed by $p^4$
(see   \EqnRef{eqn:ThyRatio}).   Matrix elements  show less dependence
upon quark mass at larger recoil momenta.  Statistical noise increases
with recoil   momenta, however, making   extrapolations there noisier.
Hence we find a ``window''  at  moderate recoil momentum where  chiral
extrapolations  are most dependable.    By  varying our  extrapolation
procedures we estimate chiral extrapolations contribute a $10\%$ error
to $T_{\pi/\ssbar}(0.7)$.   If correlators could  be reliably computed
for masses  significantly   below   $m_s$ our   chiral   extrapolation
uncertainties would  decrease.   The  Modified Quenched  Approximation
shows promise in this respect\cite{MQA}.

Errors  due to varying fitting  procedures  and residual excited state
contamination are estimated to be about $2\%$.

\section{CONCLUSIONS}

\ErrorTable

Table~\ref{tab:Errors} summarizes  sources of  errors in $\Vratio$ for
$\Pcut = 0.7\,\GeV$.  These  preliminary  results are based upon   our
$\beta=5.7$ analysis.  We are completing the analysis of $K$ and $\pi$
decays on our finer $\beta=5.9$ lattice and continuing to study cutoff
effects using our  $6.1$ lattice.  Smaller  $a$ will result in smaller
cutoff    effects than our     $\beta=5.7$  results but with   similar
statistical errors.  The error due  to the quenched approximation  has
yet to be fully investigated.  A quenching  error of $10$-$20\%$ would
not be   unexpected,  however, in  the  deviation  of $T_{\pi/K}$ from
unity.

Theoretical uncertainties in   $\Vratio$ have been  estimated  for our
procedure with $\Pcut = 0.7\,\GeV$.  We selected this cut to eliminate
the largest momentum dependent errors and chiral extrapolation errors.
Before quoting  a  value for $\Vratio$, $\Pcut$  should   be chosen to
minimize the total theoretical and experimental errors.

\vskip-4ex\hbox{}
\section*{ACKNOWLEDGMENTS}
\hbox{}\vskip-4ex
Fermilab is operated  by  Universities Research Association, Inc.  for
the U.S. Dept. of Energy.
\vskip-4ex\hbox{}

\end{document}